\documentclass[twoside,journey]{IEEEtran}
\usepackage{makecell}
\usepackage{hyperref}
\usepackage{array}
\usepackage{graphicx,amssymb,amsmath}
\usepackage{multicol}
\usepackage[noadjust]{cite}
\usepackage{setspace}
\usepackage{subfigure}
\usepackage{graphicx}
\usepackage{float}
\usepackage {url}
\usepackage{stfloats}
\usepackage{amsthm,pifont}
\usepackage{flushend}
\usepackage{cases,subeqnarray}
\usepackage{bm,multirow,bigstrut}
\usepackage{amsmath, amsthm, amssymb}
\usepackage{textcomp}
\usepackage{latexsym,bm}
\usepackage{booktabs}
\usepackage{xcolor}
\usepackage{mathtools}
\usepackage{dsfont}
\usepackage{extarrows}
\usepackage{epsfig}
\usepackage{epsfig}
\usepackage{epstopdf}
\usepackage[noend]{algpseudocode}
\usepackage{algorithmicx,algorithm}
\theoremstyle{plain}

\theoremstyle{plain}

\usepackage{amsmath}

\IEEEoverridecommandlockouts
\begin{document}

\title{Generative AI-aided Joint Training-free Secure Semantic Communications via Multi-modal Prompts}
\author{Hongyang~Du, Guangyuan~Liu, Dusit~Niyato,~\IEEEmembership{Fellow,~IEEE}, Jiayi Zhang, Jiawen Kang, Zehui Xiong, Bo~Ai,~\IEEEmembership{Fellow,~IEEE}, and Dong~In~Kim,~\IEEEmembership{Fellow,~IEEE}
\vspace{-0.08cm}
\thanks{H.~Du, G.~Liu, and D.~Niyato are with Nanyang Technological University, Singapore. J. Zhang and B. Ai are with Beijing Jiaotong University, China. J. Kang is with Guangdong University of Technology, China. Z. Xiong is with Singapore University of Technology and Design, Singapore. D.~I.~Kim is with Sungkyunkwan University, Korea.}
}
\maketitle
\vspace{-1cm}

\begin{abstract}
Semantic communication (SemCom) holds promise for reducing network resource consumption while achieving the communications goal. However, the computational overheads in jointly training semantic encoders and decoders—and the subsequent deployment in network devices—are overlooked. Recent advances in Generative artificial intelligence (GAI) offer a potential solution. The robust learning abilities of GAI models indicate that semantic decoders can reconstruct source messages using a limited amount of semantic information, e.g., prompts, without joint training with the semantic encoder. 
A notable challenge, however, is the instability introduced by GAI's diverse generation ability. This instability, evident in outputs like text-generated images, limits the direct application of GAI in scenarios demanding accurate message recovery, such as face image transmission. 
To solve the above problems, this paper proposes a GAI-aided SemCom system with multi-model prompts for accurate content decoding. 
Moreover, in response to security concerns, we introduce the application of covert communications aided by a friendly jammer. The system jointly optimizes the diffusion step, jamming, and transmitting power with the aid of the generative diffusion models, enabling successful and secure transmission of the source messages.
\end{abstract}
\IEEEpeerreviewmaketitle
\section{Introduction}
The continuous evolution of wireless communication networks has led to an exponential growth in data volume. One potential solution to this challenge is the semantic communications (SemCom) technique~\cite{yang2022semantic}. The basic architecture of SemCom involves joint training of a semantic encoder and decoder. After passing through the semantic encoder, the source message is transformed into semantic information suitable for wireless transmission. The semantic decoder can then decode this information to recover the original source message or fulfill specific task requirements.
\begin{figure}[!t]
\centering
\includegraphics[width=0.4\textwidth]{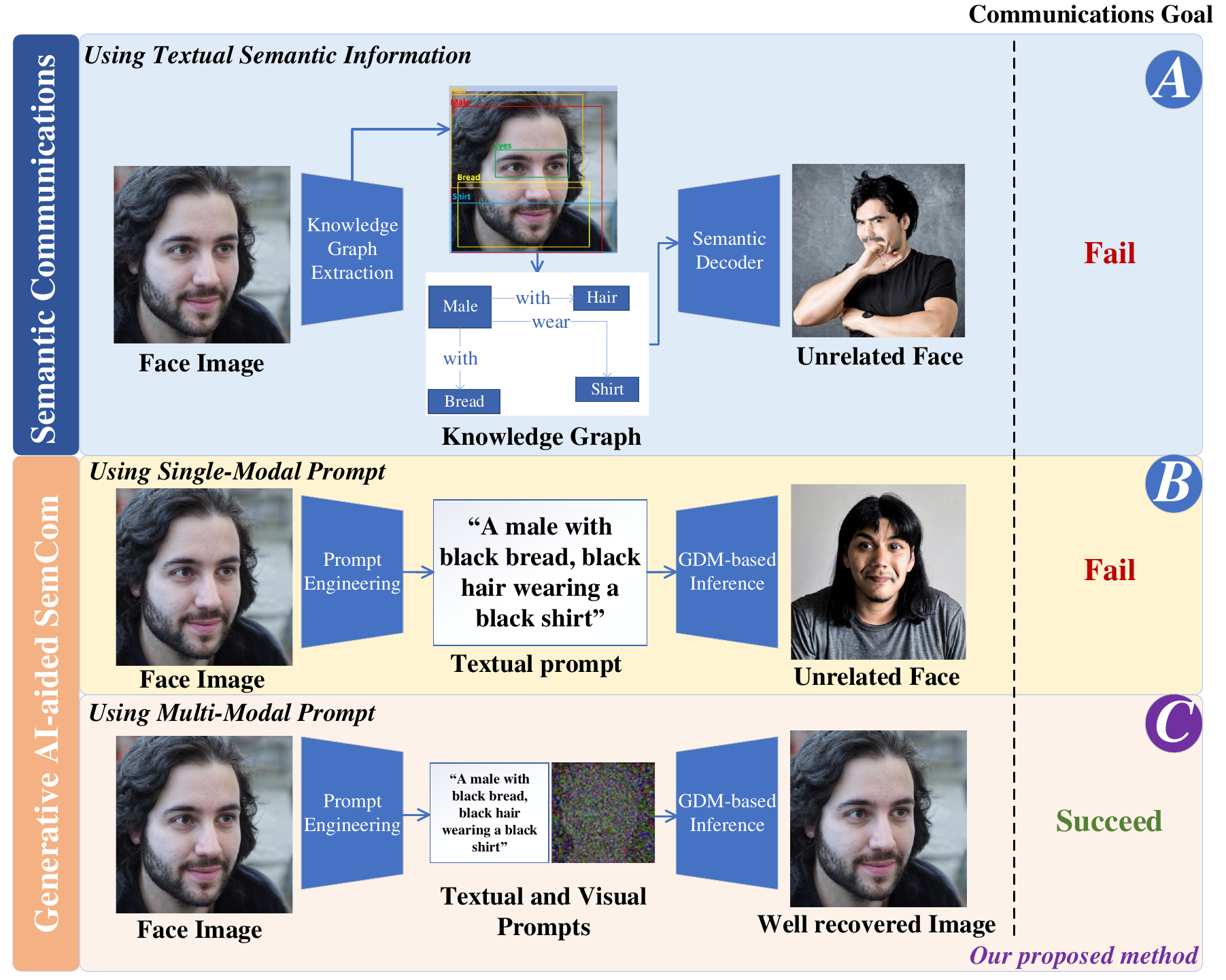}
\caption{Diverse semantic communication schemes. {\textbf{Part A}} represents SemCom based on knowledge graphs. {\textbf{Part B}} shows GAI-aided SemCom using only the textual prompt. The decoder reconstructs an image that is not accurate. {\textbf{Part C}} shows GAI-powered SemCom using multi-modal prompts. The decoder leverages GAI models to generate an accurate image.}
\label{first}
\end{figure}

Yet, there are inherent challenges in jointly training and distributing the semantic encoders and decoders, making the process both complex and energy-demanding~\cite{qin2021semantic}. Specifically, the semantic encoder-decoder pair demands co-training with the channel to achieve optimal performance~\cite{xie2021deep}. Furthermore, the trained semantic models are often tailored to specific tasks, necessitating the training and distribution of multiple semantic encoder-decoder pairs to diverse internet-of-things (IoT) devices. 
Absent this joint training, SemCom systems face difficulties in communication tasks demanding precise message reconstruction.
For example, text descriptions extracted from images, such as knowledge graphs~\cite{shi2021semantic}, can serve as the semantic information of the image. A semantic decoder can employ these graphs to extract information fit for tasks like Q\&A about image content~\cite{shi2021semantic}. Conversely, this method proves ineffectual for tasks such as transmitting face images where accuracy is paramount, as shown in Part A of Fig.~\ref{first}. Therefore, avoiding joint training while ensuring accurate image transmission is a significant challenge in SemCom.

The rise of Generative AI (GAI) introduces a potential solution to achieve the accurate transmission goal. GAI proves especially advantageous for designing decoders, enabling efficient source information retrieval without joint encoder training. In the text domain, advanced language models such as ChatGPT can craft detailed articles from simple prompts, which might serve as concise semantic information about the transmitter's message. For image, conveying a photo might involve transmitting its text description. Several image-generation GAI models, e.g., DALLE~\cite{dalle} and Stable Diffusion~\cite{stabdiff}, can perform semantic decoding after receiving prompts.
Specifically, a prompt is a succinct representation that instructs models to generate specific outputs, whether in text, images, or other digital content forms~\cite{liu2022design}. However, implementing GAI-aided SemCom poses an inherent challenge: {\textit{a single prompt can result in varied image interpretations}} as shown in Part B of Fig.~\ref{first}. This variability is attributable to the dynamic nature of GAI models~\cite{croitoru2023diffusion}. While this adaptability is advantageous in many scenarios, it becomes especially problematic for SemCom tasks that demand message accuracy, such as transmitting human face images. Hence, the design of prompts is imperative, entering the realm of prompt engineering. This paper introduces the concept of multi-modal prompts to address the challenges in SemCom tasks that require accurate image reconstruction. The multi-modal prompts incorporate {\textit{visual prompts}}, which are aimed at restoring the image's structural fidelity, and {\textit{textual prompts}}, which capture the semantic information of the image.

In addition, data security is a significant issue in the transmission process of multi-modal prompts~\cite{chen2023covert}. First, transmitting prompts within the confines of an open wireless environment necessitates robust protective measures. Covert communication is a promising and potential solution. Unlike traditional physical layer security approaches~\cite{chen2023covert}, covert communication  operates on concealing the communication activity itself, effectively making it indiscernible to potential eavesdroppers or external attackers. 
Second, visual information leakage may happen when messages are intercepted. The reason is that common visual prompts, such as the contours of objectives in the transmitted image, inadvertently lead to information leakage~\cite{liu2023semantic}. Consequently, an optimal visual prompt should not overtly reveal information about the original image. Simultaneously, it should effectively aid the GAI model in reconstructing the structure information of the original image. To address this challenge, we propose a GAI-aided secure SemCom framework as shown in Part C of Fig.~\ref{first}. The key contributions are as follows:
\begin{itemize}
\item We introduce a new GAI-aided SemCom framework without necessitating joint training. This approach offers a reduction in both computational complexity and energy cost compared to conventional SemCom methods.
\item Our novel approach leverages multi-modal prompts, allowing for the accurate reconstruction of the source message. This innovation addresses the challenge of unstable data recovery when GAI models are used.
\item We use covert communication techniques to safeguard the transmission of multi-model prompts within the open wireless environment. The optimal resource allocation scheme is generated using the generative diffusion model (GDM)-based method, achieving accurate image regeneration under energy constraints.
\end{itemize}

\section{System Model and Problem Formulation}
In this section, we discuss the system model and formulate the optimization problem.
\begin{figure}[!t]
\centering
\includegraphics[width=0.4\textwidth]{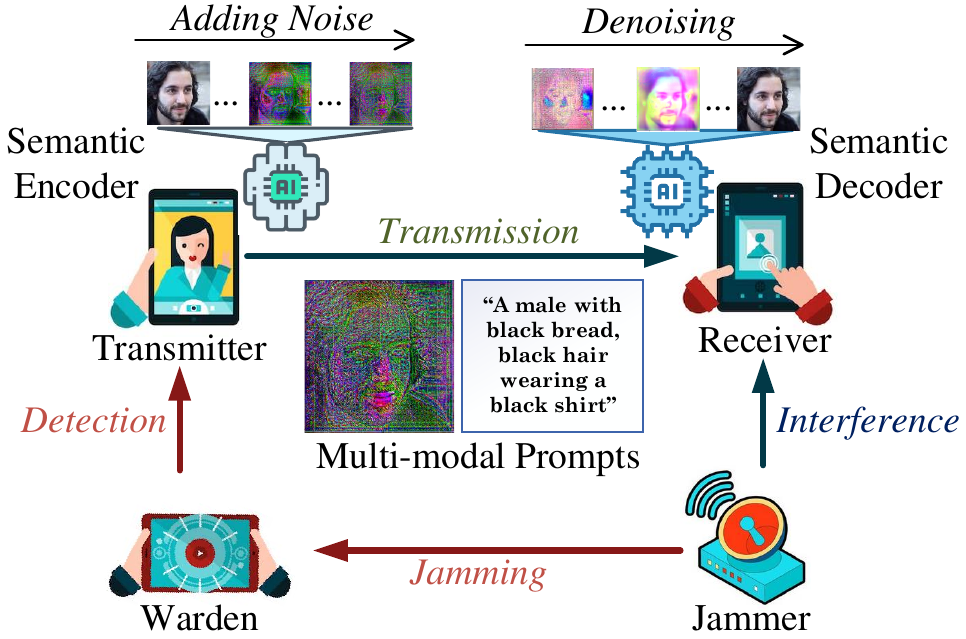}
\caption{The proposed GAI-aided secure SemCom system with the covert communications technique.}
\label{systemmodel}
\end{figure}
\subsection{Covert Communications}
The network consists of a transmitter, a receiver, a friendly jammer, and a warden in an open wireless environment. The transmitter's objective is to transmit images to the receiver while evading detection by the warden. To enhance data security, we use the covert communication technique. Instead of depending on encryption, covert communication hides the transmission behavior~\cite{chen2023covert}. Specifically, the warden evaluates two potential scenarios: the null hypothesis, $\mathcal{H}_0$, representing the transmitter's inactivity, and the alternate hypothesis, $\mathcal{H}_1$, indicating active transmission. This can be mathematically represented as:
\begin{equation}
{y_{w}} = \left\{ {\begin{array}{*{20}{l}}
{{\kappa ^2} + D_{jw}^{ - {\alpha _{jw}}}{P_{j}}h_{jw}^2}, \quad\qquad\qquad\qquad \mathcal{H}_0,	 \\ 
{D_{tw}^{ - {\alpha _{tw}}}{P_{t}}h_{tw}^2 + {\kappa ^2} + D_{jw}^{ - {\alpha _{jw}}}{P_{j}}h_{jw}^2}, \:\: \mathcal{H}_1,
\end{array}} \right.
\end{equation}
where $P_{t}$ is the transmit power, $P_{j}$ is the jamming power, ${\kappa ^2}$ characterizes the Gaussian noise. Distances between the jammer and the warden, and the transmitter and the warden are given by \( D_{jw} \) and \( D_{tw} \), respectively. \( {\alpha _{jw}} \) and \( {\alpha _{tw}} \) are the path loss exponents for their respective links, while \( h_{jw} \) and \( h_{tw} \) reflect the small-scale fading effects.

 Decision-making by the warden, denoted as $\mathcal{D}_1$ and $\mathcal{D}_0$, is grounded in the aforementioned hypotheses, adhering to a threshold rule~\cite{chen2023covert}. Detection inaccuracies occur in two situations: {\it false alarm}, where $\mathcal{D}_1$ is selected during $\mathcal{H}_0$, and {\it miss detection}, where $\mathcal{D}_0$ is chosen during $\mathcal{H}_1$. The detection error probability (DEP), which quantifies the likelihood of inaccurate warden decisions, is defined as
\begin{align}
{\xi} =& {{\mathbb P}_{FA}} + {{\mathbb P}_{MD}} = \Pr \left( {{\kappa ^2} + D_{jw}^{ - {\alpha _{jw}}}{P_{jk}}h_{jw}^2 > {\varepsilon}} \right)\notag\\&
+ \Pr \left( {D_{tw}^{ - {\alpha _{tw}}}{P_{a}}h_{tw}^2 + {\kappa ^2} + D_{jw}^{ - {\alpha _{jw}}}{P_{j}}h_{jw}^2 < {\varepsilon}} \right),
\end{align}
where $\varepsilon$ denotes the detection threshold, $\mathbb{P}_{FA}$ denotes the false alarm probability, and $\mathbb{P}_{MD}$ denotes the miss detection probability. Covert communication is successful when the DEP exceeds a threshold, i.e., $\xi_{\rm th}$, approximating $1$. 

\subsection{Problem Formulation}
We propose the GAI-aided SemCom. In this framework, a transmitter processes a source image, i.e., ${\rm Img_s}$, extracting multi-modal prompts. The prompts are transmitted over wireless channels. The receiver regenerates the image, i.e., ${\rm Img_r}$, with the GAI model. Given that diverse resource allocation strategies influence the signal-to-noise ratio (SNR), which subsequently alters the bit error probability (BEP) and impacts the final reconstruction of the image at the receiver, we employ the structural similarity (SSIM) metric as our objective function. Let $T$ represent the number of diffusion steps in the image generation process. The terms $\beta_t$, $\beta_j$, and $\beta_T$ denote the energy costs per unit for transmit power, jamming power, and diffusion step, respectively. We formulate the optimization problem as:
\begin{equation}\label{dseafa}
\begin{array}{*{20}{l}}
{\mathop {\max }\limits_{\left\{ {{P_t},{P_j},T} \right\}} }&{{\rm SSIM}\left( {{\rm Img_s},{\rm Img_r}} \right),}
\vspace{0.2cm}
\\
{\:\quad{\rm s.t.,}}&{\xi{\left(P_t,P_j\right)}  > {\xi _{\rm th}}}, \\
{}&{\beta _t}{P_t} + {\beta _j}{P_j} + {\beta_T}T \le E,
\end{array}
\end{equation}
where the first constraint ensures the communications remain covert, and the second constraint bounds the total energy $E$, which introduces a natural trade-off. On the one hand, when $P_t$ is too low or $P_j$ is too high, it becomes simpler to maintain covert communications. Yet, this configuration increases the BEP because of the low SNR, reducing the SSIM value. On the other hand, while the image generation process with high $T$ can improve the robustness against noise~\cite{du2023exploring}, excessive energy consumption towards this could limit the energy available for $P_j$ and $P_t$. As such, joint optimization is essential to balance covert communications and the quality of the regenerated image. Next, we introduce the GAI-aided SemCom in Section~\ref{afae2} and then give the optimization problem solution in Section~\ref{afae23}.

\section{Multi-modal Prompt Mechanism}\label{afae2}
In this section, we introduce the multi-modal prompt mechanism in the GAI-aided SemCom.
\subsection{Semantic Encoder}
Let us consider ${\textbf{x}}_0$ as the original image, i.e., ${\rm Img_s}$.
\subsubsection{Textual Prompt}
Textual prompts for image generation typically align with image-to-text tasks. The Blip method~\cite{li2022blip} innovatively utilizes noisy web data refined through bootstrapping. The fundamental workflow can be outlined as follows:
\begin{enumerate}
    \item \textbf{Multi-modal Mixture of Encoder-Decoder (MED):} This has three operational modes. The {\textit{Unimodal Encoder}} functions for images and text independently, using a classification token for text summarization. The {\textit{Image-grounded Text Encoder}} merges visual data with text, resulting in a multi-modal representation through an encoder token. Lastly, the {\textit{Image-grounded Text Decoder}} transforms images to text using causal self-attention and special tokens for sequence demarcation.
    
    \item \textbf{Pre-training Objectives:} MED is tailored with three key objectives. The {\textit{Image-Text Contrastive Loss}} aligns image and text features, distinguishing congruent from incongruent pairs. {\textit{Image-Text Matching Loss}} acts as a binary classifier to ascertain the alignment of visual and textual inputs. {\textit{Language Modeling Loss}} is tasked with producing coherent text from images.
    
    \item \textbf{Captioning and Filtering:} A twofold approach to manage web data noise. The {\textit{Captioner}} creates synthetic captions for online images, while the {\textit{Filter}} removes unreliable original and generated captions.
    
    \item \textbf{Data Integration:} The refined image-text combinations merge with human-annotated content, forming an exhaustive dataset for model training.
\end{enumerate}
Given an image ${\textbf{x}}_0$, the textual description ${\textbf{t}}_{\rm{sem}}$ of ${\textbf{x}}_0$ is extracted through the MED's {\textit{Image-grounded Text Decoder}} model, i.e., $\mathcal{T}\left(\cdot\right)$, as ${\textbf{t}}_{\rm{sem}} = \mathcal{T}\left\{ {\textbf{x}}_0 \right\}$.

\subsubsection{Visual Prompt}
GDMs have showcased their prowess in modeling target distributions by mastering a denoising procedure over a spectrum of noise levels~\cite{ho2020denoising}. From an arbitrary Gaussian noise map, drawn from the prior $\mathcal{N}\left({{\bf{0}}},{\bf{I}}\right)$, an ideal GDM can transform this noisy map into an image sample after $T$ denoising iterations~\cite{croitoru2023diffusion}.
A pioneering effort was made by the authors in~\cite{ho2020denoising}, who introduced a function $\varepsilon_\theta^t\left(x_t\right)$. This function ingests a noisy image ${\bf x}_t$ and predicts the corresponding noise. The GDM optimization involves the loss function $ \left| {{\varepsilon_\theta^t}\left( {{{\bf{x}}_t}} \right) - \varepsilon_a } \right| $, where $\varepsilon_a$ symbolizes the actual noise that was added to ${\bf x}_0$ to produce ${\bf x}_t$.
A significant stride in the realm of denoising is the Denoising Diffusion Implicit Model (DDIM)~\cite{song2020denoising}, which stands out due to its deterministic generative process:
\begin{equation}\label{fvgaef}
{{\bf{x}}_{t - 1}}\! = \!\sqrt {{\alpha _{t - 1}}} \left(\! {\frac{{{{\bf{x}}_t} - \sqrt {1 \!-\! {\alpha _t}} \varepsilon_\theta ^t\left( {{{\bf{x}}_t}} \right)}}{{\sqrt {{\alpha _t}} }}} \!\right) \!+\! \sqrt {1 \!-\! {\alpha _{t - 1}}} \varepsilon_\theta^t\left( {{{\bf{x}}_t}} \right),
\end{equation}
and
\begin{equation}\label{fae}
q\!\left(\mathbf{x}_{t-1}\!\!\mid \mathbf{x}_t, \mathbf{x}_0\right)\!=\!\mathcal{N}\!\left(\!\!\sqrt{\alpha_{t-1}} \mathbf{x}_0 \! + \!\sqrt{1\!-\!\alpha_{t-1}} \frac{\mathbf{x}_t\!-\!\sqrt{\alpha_t} \mathbf{x}_0}{\sqrt{1\!-\!\alpha_t}}, \mathbf{0}\!\right)\!.
\end{equation}
An intriguing aspect of DDIM is the capacity to run its generative procedure in reverse, deterministically retrieving the noise map ${{{\bf{x}}_T}}$~\cite{song2020denoising}. This map can be perceived as the latent encoding for the image ${{{\bf{x}}_0}}$. Though the reconstruction accuracy is commendable, the resultant ${{{\bf{x}}_T}}$ lacks higher-level semantics expected of a meaningful representation. This observation, combined with insights from \cite{preechakul2022diffusion}, led to the exploration of treating ${\textbf{x}}_{T}$ as a visual prompt ${\textbf{v}}_{\rm{sem}}$.

Thus, with the ${{\bf{t}}_{{\rm{sem}}}}$ to catch the high-level semantic information, the conditional DDIM can be employed to encode an image ${\textbf{x}}_0$ into the visual prompt ${\textbf{v}}_{\rm{sem}}$ to catch the image structure information~\cite{preechakul2022diffusion}, as demonstrated in~\eqref{fvgaef} as
\begin{equation}\label{faefae}
{{\bf{x}}_{t + 1}} \!= \! \sqrt {{\alpha _{t + 1}}} {{\bf{f}}_\theta }\left( {{{\bf{x}}_t},t,{{\bf{t}}_{{\rm{sem}}}}} \right) \!+\! \sqrt {1\!-\!{\alpha _{t + 1}}} {\varepsilon _\theta }\left( {{{\bf{x}}_t},t,{{\bf{t}}_{{\rm{sem}}}}} \right),
\end{equation}
where
\begin{equation}
{{\bf{f}}_\theta }\left( {{{\bf{x}}_t},t,{{\bf{t}}_{{\rm{sem}}}}} \right) = \frac{1}{{\sqrt {{\alpha _t}} }}\left( {{{\bf{x}}_t} - \sqrt {1 - {\alpha _t}} {\varepsilon_\theta }\left( {{{\bf{x}}_t},t,{{\bf{t}}_{{\rm{sem}}}}} \right)} \right).
\end{equation}
Then, an image can be regenerated accurately by using both textual and visual prompts. The visual prompt ${\textbf{v}}_{\rm{sem}}$ is defined as ${\textbf{v}}_{\rm{sem}} = \mathcal{V}_T \left\{ {{\textbf{t}}_{\rm{sem}}, {\textbf{x}}_0 } \right\}$, where $\mathcal{V}_T$ denotes the diffusion process, i.e.,~\eqref{faefae}, with $T$ steps.

\subsection{Semantic Decoder}
The purpose of the GDM-based semantic decoder is to use the textual and visual prompts, i.e., ${\textbf{t}}_{\rm{sem}}$ and ${\textbf{v}}_{\rm{sem}}$, to generate the source image ${\textbf{x}}_0$. This decoder is a conditional DDIM that models $p_\theta\left(\mathbf{x}_{t-1} \mid \mathbf{x}_t, \mathbf{t}_{\mathrm{sem}}\right)$ to match the noising distribution $ q\left( {{{\bf{x}}_{t-1}}\mid {{\bf{x}}_t},{{\bf{x}}_0}} \right) $ defined in~\eqref{fae}, with the following reverse (generative) process as:
\begin{equation}\label{flaei}
{p_\theta }\left( {{{\bf{x}}_{0:T}}\mid {{\bf{t}}_{{\rm{sem}}}}} \right) = p\left( {{{\bf{x}}_T}} \right)\prod\limits_{t = 1}^T {{p_\theta }} \left( {{{\bf{x}}_{t - 1}}\mid {{\bf{x}}_t},{{\bf{t}}_{{\rm{sem}}}}} \right),
\end{equation}
which can be further expressed as
\begin{equation}
{p_\theta }\!\left(  {{{\bf{x}}_{t - 1}}\!\mid\! {{\bf{x}}_t},{{\bf{t}}_{{\rm{sem }}}}} \right)\! =\! \left\{\!\!\! {\begin{array}{*{20}{l}}
{{\cal N}\left( {{{\bf{f}}_\theta }\left( {{{\bf{x}}_1},1,{{\bf{t}}_{{\rm{sem }}}}} \right),{\bf{0}}} \right)}&\!\!{{\rm{ if }}\: t = 1},\\
{q\left( {{{\bf{x}}_{t - 1}} \! \mid \!{{\bf{x}}_t},{{\bf{f}}_\theta }\left( {{{\bf{x}}_t},t,{{\bf{t}}_{{\rm{sem }}}}} \right)} \right)}&\!\!{{\rm{ otherwise.}}}
\end{array}} \right.
\end{equation}
Training is done by optimizing
\begin{equation}
{L_{{\rm{simple }}}} = \sum\limits_{t = 1}^T {{{\mathbb{E}}_{{{\bf{x}}_0},{\varepsilon _t}}}} \left[ {\left\| {{\varepsilon _\theta }\left( {{{\bf{x}}_t},t,{{\bf{t}}_{{\rm{sem }}}}} \right) - {\varepsilon _t}} \right\|_2^2} \right],
\end{equation}
where ${\varepsilon_t}\sim{\cal N}({\bf{0}},{\bf{I}})$ and $ {{\bf{x}}_t} = \sqrt {{\alpha _t}} {{\bf{x}}_0} + \sqrt {1 - {\alpha _t}} {\varepsilon_t}$.

\section{GDM-based Resource Allocation Scheme}\label{afae23}
In this section, we present the GDM-based resource allocation scheme for the optimization problem~\eqref{dseafa}.
\subsection{GDM in Optimization}
The application of a conditional GDM facilitates the derivation of an optimal resource allocation scheme~\cite{du2023beyond}, i.e., ${\bm r} = \left\{P_t, P_j, T\}\right.$, in \eqref{dseafa}. Distinct from traditional backpropagation techniques in neural networks or the direct model parameter optimization using deep reinforcement learning, diffusion models incrementally refine the primary distribution by denoising. We introduce vector ${\bm{c}}$ to encapsulate the multiple factors that influence the optimal resource allocation scheme, i.e., the condition, as
\begin{align}\label{env}
{\bm{c}} =  \{ & {D_{tw}},{D_{tr}},{D_{jw}},{D_{jr}},{\alpha_{tw}},{\alpha_{tr}},{\alpha_{jw}},{\alpha_{jr}}, 
\notag \\
& {\kappa ^2}, \varepsilon, \xi_{\rm th}, h_{tw}, h_{tr}, h_{jw}, h_{jr}\}.
\end{align}

\subsection{Scheme Evaluation and Generation Networks}
Introducing the \textit{scheme evaluation network}, denoted as $Q_\upsilon$, we can assign a Q-value indicative of the predicted objective function, i.e., ${\rm SSIM}\left( {{\rm Img_s},{\rm Img_r}} \right)$, to a condition-resource allocation scheme pair, i.e., ${\bm c}$ and ${\bm r}$. The $Q_\upsilon$ network serves as an informative reference for the training of the GDM-based \textit{scheme generation network} that is denoted ${\bm{\eta}}_\theta$. The ideal ${\bm{\eta}}_\theta$ is designed to generate a resource allocation scheme ${\bm r}_0$ with the maximal predicted Q-value by progressive denoising with ${\bm{c}}$ being the condition, commencing from Gaussian noise like \eqref{flaei}. However, the training loss function differs, which can be represented as:
\begin{equation}\label{actortrain}
\mathop {\arg \min }\limits_{{{\bm{{\bm{\eta}}}}_\theta }} \mathcal{L}_{{\bm{\eta}}}(\theta) = - {\mathbb{E}_{{{\bm{r}}_0}\sim{{\bm{\eta}}_\theta }}}\left[ {{Q_\upsilon }\left( {{\bm{c}},{{\bm{r}}_0}} \right)} \right].
\end{equation}

The \textit{solution evaluation network}, $Q_\upsilon$, aims to minimize the difference between predicted and actual Q-values. Therefore, the loss function for $Q_\upsilon$ can be formulated as:
\begin{equation}\label{qualitytrain}
\mathop {\arg \min }\limits_{{Q_\upsilon }} \mathcal{L}_Q(\upsilon) = {\mathbb{E}_{{\bm{r}}_0\sim{{\bm{\eta}} _{{\theta }}}}} \left[ {{{\left\| { {r}\!\left( {{\bm{c}},{{\bm{r}}_0}} \right)  -  {Q_{{\upsilon}}}\left( {{\bm{c}},{{\bm{r}}_0}} \right)} \right\|}^2}} \right],
\end{equation}
where the \(r\) is the real objective value, ${\rm SSIM}\left( {{\rm Img_s},{\rm Img_r}} \right)$, obtained when the resource allocation scheme ${{\bm{r}}_0}$ is implemented under the condition ${\bm{c}}$.

\section{Numerical Analysis}
\begin{figure*}[t]
\centering
\begin{minipage}[t]{0.3\textwidth}
\centering
\includegraphics[width=0.98\textwidth]{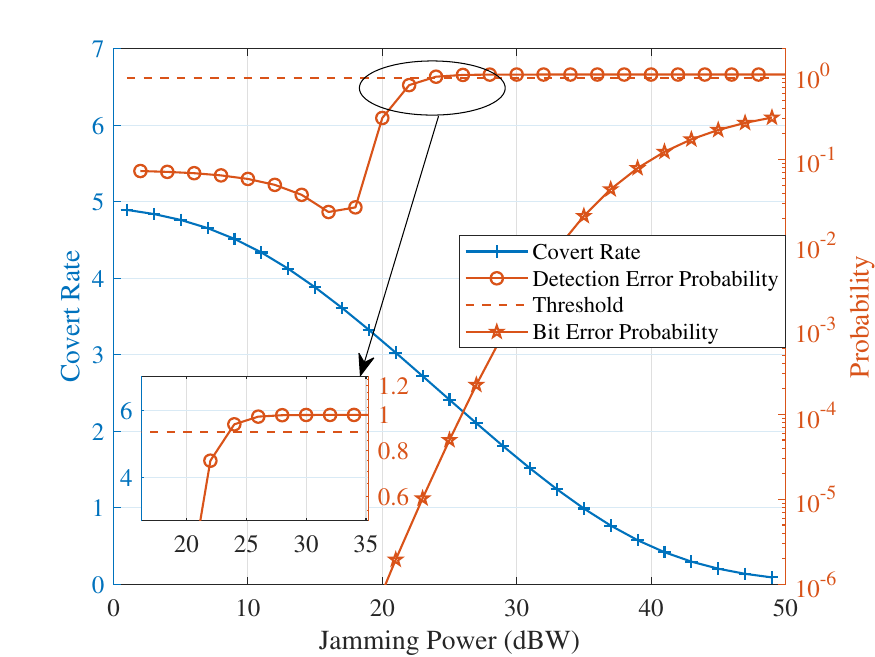}
\caption{The covert rate, DEP, and the BEP versus the jamming power.}
\label{res}
\end{minipage}
\hspace{0.1cm}
\begin{minipage}[t]{0.3\textwidth}
\centering
\includegraphics[width=0.98\textwidth]{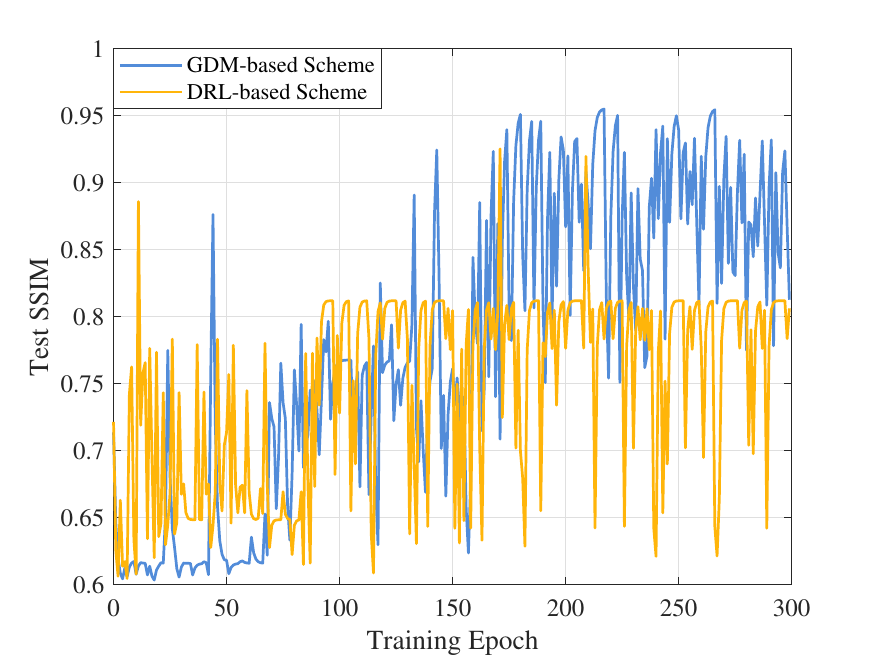}
\caption{Test rewards of GDM-based and DRL-based resource allocation schemes when the learning rates of ${\bm{\eta}}_\theta$ and $Q_\upsilon$ networks are $10^{-4}$.}
\label{img1}
\end{minipage}
\hspace{0.1cm}
\begin{minipage}[t]{0.3\textwidth}
\centering
\includegraphics[width=0.85\textwidth]{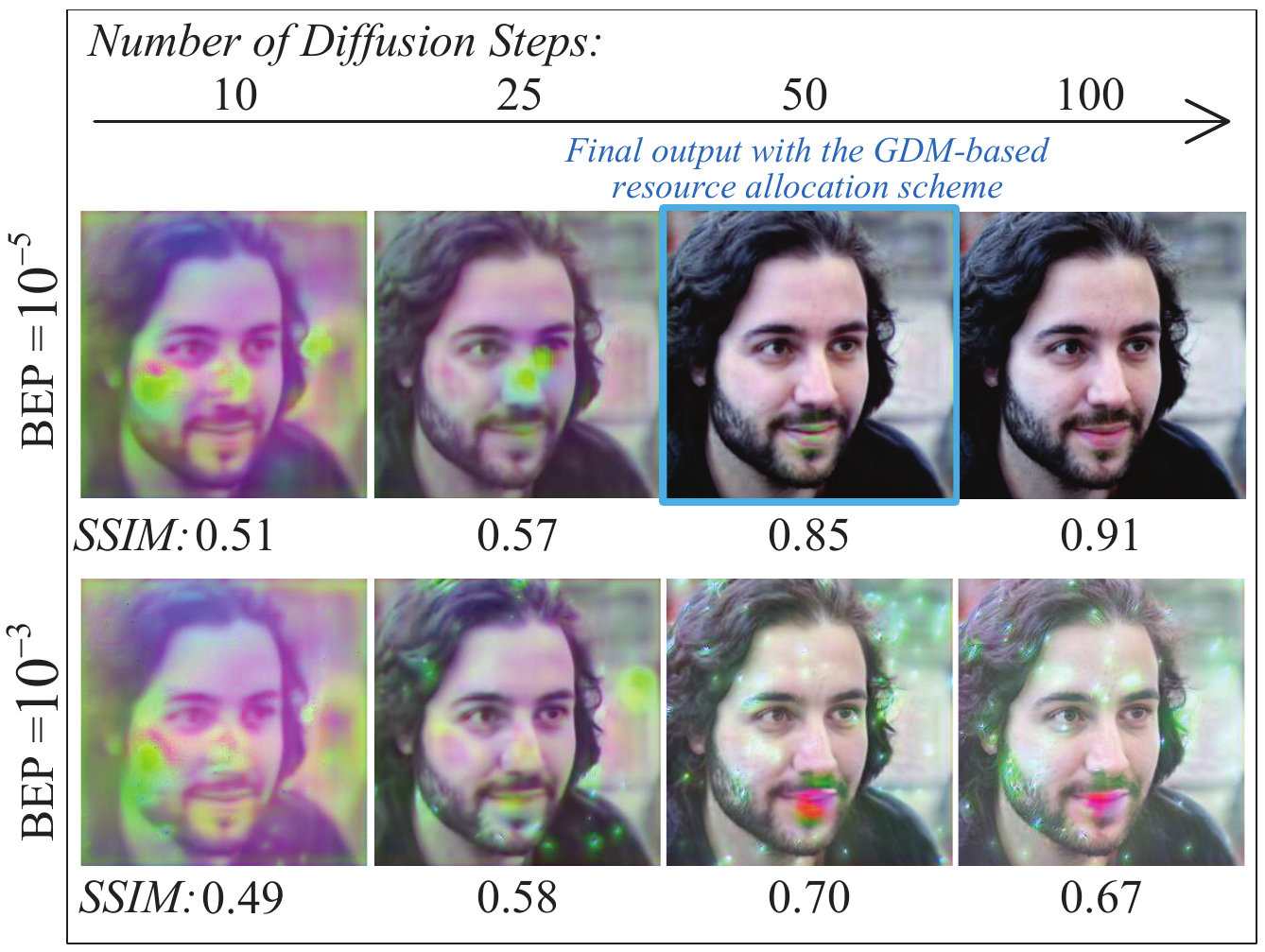}
\caption{The regeneration process when the receiver uses the received multi-modal prompts with different BEP and diffusion steps.}
\label{img2}
\end{minipage}
\end{figure*} 
In this section, we demonstrate the feasibility of the proposed GAI-aided secure SemCom system and the effectiveness of the proposed GDM-based resource allocation scheme.

In Fig.~\ref{res}, the impact of increased jamming power on the covert rate (defined as the data rate achieved during covert communications), DEP, and BEP are depicted, with $P_t = 20$ ${\rm dBW}$, $\varepsilon = 50$. Positioned within a Cartesian coordinate system with units in meters, the transmitter, warden, receiver, and jammer have coordinates at $(3,8)$, $(3,14)$, $(7,10)$, and $(6,8)$, respectively. The path loss exponents are $\alpha_{tr} = 1$, $\alpha_{tw} = 1.2$, $\alpha_{jw} = \alpha_{jr} = 1.7$. Small-scale channel fading attributes, ${\rm dBW}$, $h_{tw}$, $h_{tr}$, $h_{jw}$, and $h_{jr}$, are in line with the $\alpha$-$\mu$ fading model, with parameters $\alpha=2$ and $\mu=4$. The channel coding approach adopted is Binary Phase-shift Keying (BPSK). 
From Fig.~\ref{res}, we can observe that as the jamming power increases, there is a consistent decrease in the covert rate and an increase in the BEP. Given the current environmental conditions and the warden's detection threshold, covert communication requirements are satisfied, which means that the DEP exceeds the threshold when the jamming power exceeds approximately $24$ $\text{dBW}$. However, this results in a BEP greater than or equal to $10^{-5}$.

The test reward curves of the GDM-based and deep reinforcement learning (DRL)-based resource allocation schemes are presented in Fig.~\ref{img1}. The well-trained resource allocation scheme generation network ${\bm{\eta}}_\theta$ determines the number of diffusion steps, transmit power, and jammer power values to enable covert communications under the given condition ${\bf c}$, resulting in a BEP. Subsequently, image regeneration is performed. It is shown that the GDM method overperforms the DRL method. Fig.~\ref{img2} shows the regeneration process using the received multi-modal prompts with different diffusion steps under varying BEPs. From Fig.~\ref{img2}, we can observe that when BEP is $10^{-5}$, approximately $50$ diffusion steps are sufficient to achieve a reasonable image reconstruction quality. However, as the BEP degrades to $10^{-3}$, the reconstructed images consistently exhibit noise, leading to relatively lower SSIM scores. This further underscores the essentiality of a well-designed resource allocation scheme.

\section{Conclusion}
We presented a GAI-aided secure SemCom system to address challenges in computational overhead and data security in semantic communication. By eliminating joint training and employing multi-modal prompts, our approach ensures accurate message reconstruction. With the integration of covert communication techniques, our system enhances the secure transmission of prompts, offering improvements for wireless communication scenarios.

\bibliographystyle{IEEEtran}
\bibliography{Ref}

\end{document}